\documentclass[11pt]{article}

\def\gappeq{\mathrel{\rlap {\raise.5ex\hbox{$>$}} {\lower.5ex\hbox{$\sim$}}}}
\def\lappeq{\mathrel{\rlap{\raise.5ex\hbox{$<$}} {\lower.5ex\hbox{$\sim$}}}}

\def\beq{\begin{equation}} \def\eeq{\end{equation}} \def\beqa{\begin{eqnarray}}
\def\eeqa{\end{eqnarray}}\def\bq{\begin{quote}} \def\eq{\end{quote}}

\def\nn{\nonumber}

\begin{document} 

\pagestyle{empty}

\vskip 1.5 cm    
\def\thefootnote{\fnsymbol{footnote}}
\begin{center}  
{\large \bf Product Groups, Discrete Symmetries,
and Grand Unification \protect\footnote{Proceedings
of SUSY02, 10th International Conference on {\it Supersymmetry and 
Unification of Fundamental
Interactions}, 17-23/06/'02, DESY, Hamburg, Germany.}}     
\end{center}  
\vspace*{5mm} 
\centerline{\bf Yael Shadmi}
\vskip 0.5 cm
\centerline{\em Physics Department, Technion}
\centerline{\em Haifa, 32000, Israel} 
\vskip 1.5 cm   

\centerline{\bf Abstract}  
We present GUT models based on an $SU(5)\times SU(5)$ 
GUT group. These models maintain the main successes
of simple-group GUTs but permit simple solutions to
the doublet-triplet splitting problem.
Moreover, GUT breaking is triggered by supersymmetry breaking
so that the GUT scale is naturally generated as a combination
of the Planck scale and the supersymmetry breaking scale. 
\vspace*{1cm} 
\pagestyle{plain}


\def\thefootnote{\arabic{footnote}}
\setcounter{footnote}{0}

\section{Introduction and summary}
The unification of couplings in the MSSM is often viewed
as a hint for a grand unified theory (GUT) as well as for supersymmetry.
Here we will explore the possibility that the relation between
supersymmetry and grand unification is even deeper, and
that GUT breaking is a result of supersymmetry 
breaking~\cite{othergut}.
The GUT breaking VEV in our models corresponds to an approximately
flat direction, which is only lifted by higher dimension
superpotential terms, suppressed by $M_{\rm Planck}$.
Once supersymmetry is broken, if some of the GUT-breaking fields  
obtain negative soft masses$^2$ around the TeV, the GUT is broken,
and the GUT scale is determined as a combination of the Planck
scale and the TeV.
How does one generate an almost flat potential for the
GUT breaking fields? An obvious way is to charge these fields
under some discrete symmetry. Such a symmetry can forbid all
superpotential terms up to some desired order.

Discrete symmetries can also lead to doublet-triplet splitting.
Witten recently revisited this problem\cite{witten}, and beautifully
explained why it is readily solved in product-group GUTs\cite{pgroups,
Barr:1996kp}.
The idea is to impose a (global) discrete symmetry which
is broken at the GUT scale.
Below the GUT scale, a combination of this symmetry and the
GUT symmetry remains unbroken and forbids the doublet mass.
If the GUT group is simple, a discrete symmetry cannot distinguish
between the triplet and doublet mass term.
In a semi-simple GUT however, with Higgses transforming
under different group factors, the doublet mass term
and triplet mass term can have different charges under
the unbroken discrete symmetry.
We will explain this mechanism in detail in Section~\ref{doublettriplet}.

As it turns out, the product group structure\cite{kramer}, together
with the discrete symmetry, is crucial both for doublet-triplet
splitting, and for ensuring an approximately flat potential
for the GUT breaking fields.

Furthermore, in a product group GUT, the standard model matter 
fields and Higgses may originate  from different GUT group factors.
This has a number of consequences, which are all related to the
fact that some of the Yukawas will appear as non-renormalizable
terms, and will therefore be suppressed by some power of 
$M_{\rm GUT}/M_{\rm Pl}$.
Clearly one gets some non-trivial hierarchies of Yukawa couplings.
Second, the Higgs-triplet couplings to standard model fields
can be suppressed relative to the usual Yukawa couplings.
Thus the proton decay constraint on the triplet mass can
in principle be relaxed.
Proton decay is then suppressed by a combination of doublet-triplet
``mass splitting'' (the usual mechanism) and ``Yukawa splitting''.
Third, matter fields of different generations may transform
under different GUT group factors, so that bottom-tau mass unification
can still hold, while similar relations for the first generations
are avoided.

Product group GUTs often seem problematic, because
the standard model gauge couplings arise from different
couplings at the high scale. To maintain unification some
GUT group factors have to be strongly coupled.
In addition, hypercharge quantization is lost.
None of this happens in the models we consider.
Because the standard model is a subgroup of the diagonal $SU(5)$,
hypercharge arises in the usual way, and the three
couplings originate from a single coupling---the coupling of
the diagonal group.

\section{A model}
We will now present an explicit model and use it to demonstrate
the ideas we discussed. The gauge group is $SU(5)\times SU(5)$,
and we impose a discrete symmetry $Z_N\times Z_M^{R}$, the
latter being an $R$-symmetry.
The fields and their charges are summarized in Table~\ref{tab:approx}.
\begin{table}[ht]
\begin{center}
\begin{tabular}{cc} \hline \hline
Field           &  $SU(5)\times SU(5)\times Z_N\times Z_M^R$ \\ \hline\hline
$\Phi_1$        &  $(5,\bar5,1,0)$      \\
$\bar\Phi_1$    &  $(\bar5,5,N-1,0)$    \\
$\Phi_2$        &  $(5,\bar5,(N-3)/2,0)$        \\
$\bar\Phi_2$    &  $(\bar5,5,(N+3)/2,0)$        \\
$A_1$           &  $(24,1,(N-5)/2,1)$ \\
$A_2$           &  $(24,1,(N+5)/2,1)$ \\
$A_3$           &  $(24,1,0,1)$ \\
$S$             &  $(1,1,0,M-1)$ \\ \hline
$h$             &  $(5,1,1,1)$          \\
$\bar h^\prime$ &  $(1,\bar5,0,0)$      \\ \hline
$\bar h$        &  $(\bar5,1,0,0)$      \\
$h^\prime$      &  $(1,5,N-1,1)$                \\
\hline\hline
\end{tabular}
\end{center}
\caption{The fields of the basic model}
\label{tab:approx}
\end{table}
The symmetries allow the following superpotential
\beqa\label{typicalw}
W&=&\lambda_{12}\Phi_1 A_1\bar\Phi_2+\lambda_{21}\Phi_2 A_2\bar\Phi_1+
\lambda_{11}\Phi_1 A_3\bar\Phi_1+\lambda_{22}\Phi_2 A_3\bar\Phi_2
+\eta_{12}SA_1A_2+\eta_{33}SA_3A_3 \nn \\
&+& {1\over M_{\rm Pl}^{M-4}}S^{M-1}+ 
{1\over M_{\rm Pl}^{M-2}}S^{M-1} \Phi_i\bar\Phi_i +\cdots
\eeqa
where we do not show non-renormalizable terms that do not contribute to
the scalar potential, and the dots stand for higher-dimension terms.

We will be interested in the direction
\beqa\label{vev}
\langle\Phi_1\rangle=\langle\bar\Phi_1\rangle&=&
v_1\times{\rm diag}(1,1,1,0,0),\nn\\
\langle\Phi_2\rangle=\langle\bar\Phi_2\rangle&=&
v_2\times{\rm diag}(0,0,0,1,1),\nn\\
\langle S\rangle&=&s,\\
\langle A_i\rangle\nn&=&0. \eeqa 
Along this direction, it is easy to see that with 
$\lambda_{11}v_1^2=\lambda_{22}v_2^2$,
the only contribution to the scalar potential is from
the non-renormalizable terms appearing in~(\ref{typicalw}).
The bi-fundamental VEVs indeed break $SU(5)\times SU(5)$
down to the standard model gauge group.
The three adjoints $A_i$ and the singlet are required in
order to give mass to all the Goldstone bosons.

\subsection{Splitting doublets from triplets}\label{doublettriplet}
Let us first discuss how doublet-triplet splitting works here.
The bi-fundamental VEVs of~(\ref{vev}) preserve a discrete symmetry
$Z_N^\prime$ which is a combination of the original $Z_N$
and a discrete subgroup of hypercharge of, say, the first $SU(5)$.
(The latter is of the form 
$${\rm diag}(\alpha^{-1},\alpha^{-1},\alpha^{-1},
\alpha^{(N+3)/2},\alpha^{(N+3)/2})$$
for $N$ even.)
This unbroken symmetry distinguishes between the Higgs doublets
and triplets. Furthermore, if the Higgses are charged under
{\it different} $SU(5)$ factors, their mass terms have different
$Z_N^\prime$ charges. Thus, the doublet mass term can be
forbidden while the triplet mass term is allowed.
This is not the case if the Higgses are a $5$ and a $\bar 5$ of the
same $SU(5)$. We then see why this idea only works in the
context of semi-simple GUTs.

Imagine then, that the standard model Higgses are the fields
$h$ and $\bar h^\prime$ of Table~\ref{tab:approx}.
We may want to add the remaining fields $h^\prime$ and $\bar h$
to cancel anomalies (we will discuss different possibilities
shortly).
The most general renormalizable 
superpotential that involves
the $h$ fields is:
\beq\label{suupoh}
W_1=h\bar\Phi_1\bar h^\prime+h^\prime\Phi_1\bar h.
\eeq
Since the $h$ fields do not couple to the $\Phi_2$ and $S$ fields, only the
triplets acquire masses.

\subsection{GUT breaking and supersymmetry breaking}
The symmetries of our model can ensure an almost flat potential
for the GUT fields. Roughly, the potential scales as
\beq\label{scale}
V\sim {v^{2n-2}\over M_{\rm Pl}^{2n-6}}\ ,  
\eeq
where $v$ is the typical VEV and in our example $n$ is either
$M-1$ or $M+1$ ($M$ is associated with the discrete symmetry $Z_M^R$
as we will now see, we will want to take $M\sim10$, so terms
with $n=M+1$ of $n=M-1$ will give similar results).
If some or all of the GUT breaking fields get a negative soft mass-squared
$m^2<0$ of order the weak scale, the minimum of the potential is
at 
\beq\label{gut}
v\sim \left( {m\over M_{\rm Pl}}\right)^{{1\over n-2}}\, M_{\rm Pl}\ .
\eeq
This is around 10$^{16}$GeV for $n\sim10$.

\section{Standard model Higgses and Yukawa couplings}
So far, we concentrated on the GUT breaking fields.
We saw that we can break the GUT symmetry to the standard
model gauge group, and give mass to all GUT breaking fields.
We could also generate the GUT scale from supersymmetry breaking.
Finally, the discrete symmetry of the model allowed mass terms for 
all Higgs triplets. 
The unbroken discrete symmetry further forbids mass terms for
all Higgs doublets. Thus, if we have two pairs of $5$ and $\bar5$
Higgses as in Table~\ref{tab:approx}, we are left with four light
doublets.
 
There are then three options:
\begin{itemize}
\item {\bf A}: The theory does not contain $h^\prime$ and $\bar h$.
$SU(5)$ anomalies are cancelled by appropriate choices of $SU(5)\times SU(5)$
representations for the standard model matter fields.
The standard model Higgses come from fields charged under different $SU(5)$'s.
Some standard model Yukawa couplings arise from non-renormalizable terms.
$Z_N^\prime$ can be broken by supersymmetry breaking effects
to generate the $\mu$ term.
\item {\bf B}: The theory does contain $h^\prime$ and $\bar h$,
but these remain massless\footnote{It is easy to forbid
the relevant mass terms by choosing appropriate charges
for  $h^\prime$ and $\bar h$.}~\cite{witten}.
Witten speculates that these could be the messengers
of supersymmetry breaking.
The heavy triplets are from $h$ and $\bar h^\prime$.
\begin{itemize}
\item {\bf B1}: The standard model Higgses come from fields charged under
different $SU(5)$'s, say, $h$ and $\bar h^\prime$, so that, again,
some standard model Yukawa couplings arise from non-renormalizable terms.
\item {\bf B2}: The standard model Higgses come from fields charged under
a single $SU(5)$, say, $h$ and $\bar h$. Then, standard model fields can all
be charged under the same $SU(5)$, and all Yukawa couplings are renormalizable.
\end{itemize}
\item {\bf C}: The theory does contain $h^\prime$ and $\bar h$.
All triplets gain mass through the couplings~(\ref{suupoh}).
The $Z_N^\prime$ is broken at a high scale,  so that one
doublet pair also gets mass around $M_{\rm GUT}$.
It is possible to arrange for an acceptable $\mu$
term for the remaining two doublets, for example, through the
mechanisms proposed in \cite{Leurer:1993gy} or in \cite{Barr:1996kp}.
This is most easily done by adding a gauge-singlet, $S_H$,
charged under the $Z_N$, with a GUT scale VEV. Then, in order to have
masses of order $M_{\rm GUT}$,
the doublets that couple to $S_H$ must be charged under the
same $SU(5)$, and the Higgs doublets are charged under the second $SU(5)$.
\end{itemize}

This clearly leads to interesting predictions for quark and lepton
masses. Some of the Yukawa terms can only arise from
higher-dimension operators involving powers of the bi-fundamental
fields and suppressed by $M_{\rm Pl}$.
In addition, because of the discrete symmetry, when some
doublet Yukawa couplings are allowed, the corresponding
triplet coupling is forbidden. This gives rise to
triplet ``Yukawa splitting''. For further details, we refer
the reader to reference~\cite{us}.
  
\vskip 0.5 cm
\noindent {\bf Acknowledgements} 
I would like to thank Michael Dine and Yossi Nir for
their collaboration on this work. 
Research supported in part by a Technion Management Career Development
Chair, and by Technion research grant \#090-018.




\end{document}